\documentclass[conference, 10pt]{IEEEtran}
\IEEEoverridecommandlockouts
\usepackage{cite}
\usepackage{amsmath,amssymb,amsfonts}
\usepackage{graphicx}
\usepackage{textcomp}
\usepackage{xcolor}
\usepackage{subcaption}
\usepackage{bm}
\usepackage{multirow}
\usepackage{booktabs}
\usepackage{subcaption}
\usepackage{amsthm}\usepackage{balance}\usepackage{url}
\usepackage{algorithm,algpseudocode}
\usepackage{amssymb}
\usepackage{eqnarray}

\newtheorem*{proof*}{Proof}

\algrenewcommand\algorithmicrequire{\textbf{Initialization:}}
\algrenewcommand\algorithmicensure{\textbf{Output:}}
\algdef{SE}[DOWHILE]{Do}{doWhile}{\algorithmicdo}[1]{\algorithmicwhile\ #1}%

\makeatletter

\newcommand{\Rmnum}[1]{\expandafter\@slowromancap\romannumeral #1@}
\makeatother

\date{}

\def\BibTeX{{\rm B\kern-.05em{\sc i\kern-.025em b}\kern-.08em
    T\kern-.1667em\lower.7ex\hbox{E}\kern-.125emX}}

\begin{document}
\title{Toward Reliable Semantic Communication: Beyond Average Performance
}

\author{Boyuan Li, 
    Mingze Gong,
     Shuoyao Wang,~\IEEEmembership{Senior Member,~IEEE}, Jia Yan~\IEEEmembership{Member,~IEEE}, \\
     Suzhi Bi,~\IEEEmembership{Senior Member,~IEEE}, and Ying-Jun Angela Zhang,~\IEEEmembership{Fellow,~IEEE}\thanks{B. Li, M. Gong, S. Wang, and S. Bi are with College of Electronic and Information Engineering, Shenzhen University, China. J. Yan is with Intelligent Transportation Thrust, The Hong Kong University of Science and Technology (Guangzhou), China. Y.~J.~A.~Zhang is with Information Engineering, The Chinese University of Hong Kong, Hong Kong.}
    }

\maketitle

\begin{abstract}
Semantic communication has emerged as a promising paradigm for improving transmission efficiency by conveying task-relevant semantics rather than raw data. Although recent studies have achieved notable gains in communication efficiency and average task performance, reliability remains a fundamental bottleneck in dynamic and uncertain environments. In particular, most existing designs are still optimized mainly for average-case behavior, while \textit{lower-tail performance} under adverse transmission conditions remains insufficiently understood and inadequately protected. In this article, we present a unified perspective on reliable semantic communication \textit{beyond average performance}. We first review three  reliability-oriented design categories: channel-aware adaptation, robustness-oriented codec design, and hybrid automatic repeat request (HARQ)-based retransmission. We show that these approaches address reliability from complementary perspectives, but each still has inherent limitations. 
Adaptation-based methods enhance performance across varying channel states through CSI-conditioned semantic coding, but they remain vulnerable to CSI mismatch and generally optimize average quality within each estimated condition rather than lower-tail reliability;
robustness-based methods improve intrinsic codec resilience but may introduce a tradeoff between average performance and stability; and HARQ-based methods exploit receiver-side sample-wise feedback but require additional communication overhead and round-trip latency. Motivated by these observations, we discuss two  solution directions: robust adaptive semantic communication under imperfect CSI, and joint source-channel-check coding with adaptive retransmission for sample-level reliability enhancement. Finally, we outline several future research directions, including the joint design of robustness and retransmission, reliability metrics beyond averages, and compatibility with existing digital wireless networks.
\end{abstract}


\section{Introduction}
\begin{figure*}[!t]
    \centering
    \includegraphics[scale=0.65 ,trim=5cm 5cm 5cm 5cm, clip]{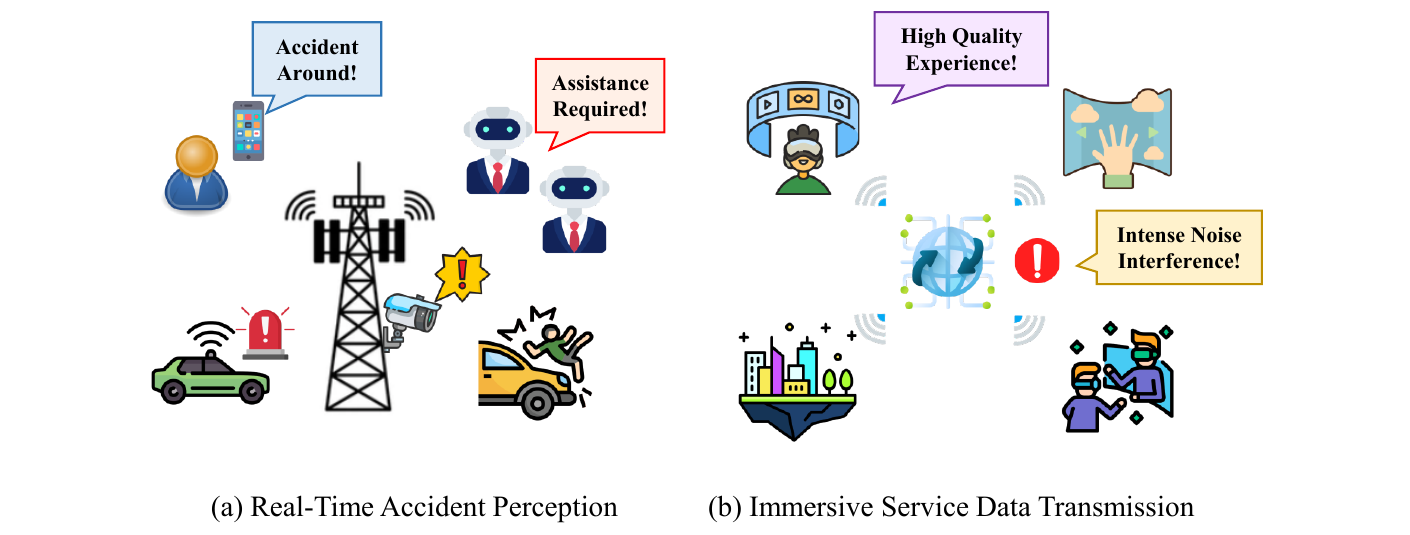}
    \caption{Example application scenarios of reliable semantic communication system.}
    \label{fig:condition}
\end{figure*}

The rapid evolution of sixth-generation (6G) wireless networks, coupled with the proliferation of ultra-high-definition multimedia services, is fundamentally reshaping the requirements for wireless communication systems.
Emerging applications, such as extended reality and metaverse-driven immersive interaction, demand not only massive data delivery but also stringent perceptual fidelity.
As a result, conventional communication paradigms are increasingly challenged by the explosive data volume as well as the stringent requirements for ultra-reliable transmission\cite{6GSurvery}.
Semantic communication (SemComm) has recently emerged as a promising paradigm to address these challenges. By extracting and transmitting high-level semantic features rather than raw data, it offers a new path to improving task-oriented communication efficiency beyond the design philosophy of traditional Shannon-style communication systems.
More importantly, by aligning the decoding objective with task-level fidelity, it inherently enhances the robustness of downstream tasks against channel impairments.
Owing to these advantages, it has attracted substantial research attention as a key enabler for next-generation intelligent communication systems.  {However, the shift from bit accuracy to semantic fidelity introduces new challenges in system design. In particular, ensuring high-quality semantic information at the receiver requires carefully coordinated encoding and transmission strategies, which must account for both channel impairments and task-specific requirements. As a result, achieving reliable semantic communication, in terms of consistently preserving semantic integrity under practical wireless conditions, remains an open and critical research problem.} Recent studies have extensively explored reliability-oriented SemComm, particularly leveraging deep learning-based joint source-channel coding (JSCC) architectures, as shown in Fig.~\ref{fig:condition}. By optimizing encoder-decoder structures in an end-to-end manner, these approaches have demonstrated notable gains in average task performance. 

Despite its potential, most existing works primarily focus on improving average-case performance, whereas reliability under highly time-varying channels, i.e., the lower-tail performance of the system, remains insufficiently addressed. In practice, a SemComm system may achieve satisfactory average reconstruction or task quality under nominal conditions, while a small fraction of transmissions may still suffer severe semantic distortion due to instantaneous fading, noise bursts, or CSI mismatch. Such lower-tail samples contribute little to mean performance metrics, but they can dominate user-perceived failures and compromise system reliability. This issue is particularly critical in highly dynamic and harsh environments, such as deep-space exploration, underwater imaging, and disaster monitoring, where severe channel perturbations, limited resources, and strong uncertainty jointly impose stringent reliability requirements.


To address these challenges, existing SemComm frameworks must evolve beyond conventional designs and incorporate three essential capabilities:
\textit{channel-aware adaptive semantic extraction, robust feature encoding mechanism, and task-oriented reliable transmission strategy.}

From the perspective of semantic extraction, dynamic wireless environments introduce significant variability in channel states. Conventional semantic encoders, typically designed for fixed Channel State Information (CSI), like Signal-to-Noise Ratio (SNR), often fail to generalize to such time-varying scenarios.
Adaptive semantic extraction, which dynamically adjusts feature representation according to channel conditions and bandwidth restriction, provides a promising direction to enhance system adaptability and improve performance across diverse communication states.

However, most existing adaptive schemes rely on the assumption of accurate CSI. In practice, CSI is often imperfect or partially observable, especially in fast-fading or highly uncertain environments. Under such conditions, overly specialized feature extraction may become fragile and susceptible to random noise interference, leading to significant degradation in task performance.
 {By learning channel-agnostic semantic representation, robustness-oriented approaches aim to mitigate performance degradation caused by CSI mismatch. Through a unified end-to-end training framework across diverse channel conditions, these methods enable the system to maintain stable and resilient performance under varying noise and uncertainty.}
Such designs improve decoding reliability under CSI estimation errors and stochastic interference, effectively mitigating noise-induced distortions and ensuring stable semantic representation under uncertainty.

Meanwhile, heterogeneous multimedia services exhibit diverse requirements in terms of efficiency and reliability. Such diversity motivates the need for a task-oriented reliable transmission strategy that goes beyond traditional encoding robustness. In particular, hybrid automatic repeat request (HARQ)-based strategies can be extended to SemComm by incorporating task-aware error detection and feature-level distortion estimation. By selectively retransmitting severely corrupted semantic features, these approaches enable precise error compensation at the receiver, thereby effectively improving the system performance lower bound. Instead of solely enhancing encoding robustness, HARQ-based strategies further leverage received feature quality to achieve more accurate  {error-detection and efficient error-correction, achieving higher transmission reliability}. However, the retransmission efficiency on error-correction is largely confined by estimation precision. Moreover, under stringent communication overhead, how to fully balance transmission reliability and efficiency, is still an open problem. 

Building upon the above analysis, adaptation provides a necessary starting point for reliable semantic transmission, because it enables channel-aware semantic encoding under varying wireless conditions. However, such adaptation mainly improves performance across different estimated channel states. For a given channel state, individual samples may still suffer severe distortion due to instantaneous noise, CSI mismatch, or random perturbations. Therefore, adaptation alone is insufficient to ensure reliable SemComm under practical uncertainty. Robustness-oriented designs are further needed to maintain stable codec behavior under imperfect CSI, while HARQ-based mechanisms are required to perceive and recover severely corrupted samples at the receiver side. In this way, reliable SemComm should jointly consider channel-aware adaptation, robustness-oriented encoding, and feedback-driven retransmission, so that both long-term stability and sample-level reliability can be improved.


The contributions of this article are three-fold. First, we identify the fundamental tension between channel-adaptive semantic coding and robustness to channel state information (CSI) imperfections, and develop a robust adaptive semantic communication framework that maintains reliable performance under practical imperfect CSI scenarios. Second, to cope with the separate module design for feature coding and check coding, we develop a joint source-channel-check coding transmission mechanism, where check codewords provide complementary semantic information and retransmission decisions are adaptively optimized to improve sample-level reliability. 
Third, we provide a unified perspective on reliable semantic communication \textit{beyond average performance}, systematically analyzing adaptation-based, robustness-based, and HARQ-based approaches, and identifying their complementary roles and remaining gaps to guide future research directions.

This article is organized as follows. We focus on the design of reliable SemComm systems and develop a unified framework that jointly considers semantic extraction, feature encoding, and transmission strategy. Specifically, we first establish a systematic perspective on reliability in SemComm by decomposing it into the aforementioned three key components, enabling adaptability to diverse tasks and service requirements. Building upon this framework, we then present two potential solution directions on semantic extraction, feature encoding and transmission strategy, all of which demonstrate significant performance gains in non-ideal transmission environments. Finally, we discuss open research challenges and outline promising future directions toward fully reliable SemComm systems.

\begin{figure*}[!t]
    \centering
    \includegraphics[scale=0.52 ,trim=5cm 0.5cm 0cm 0.5cm, clip]{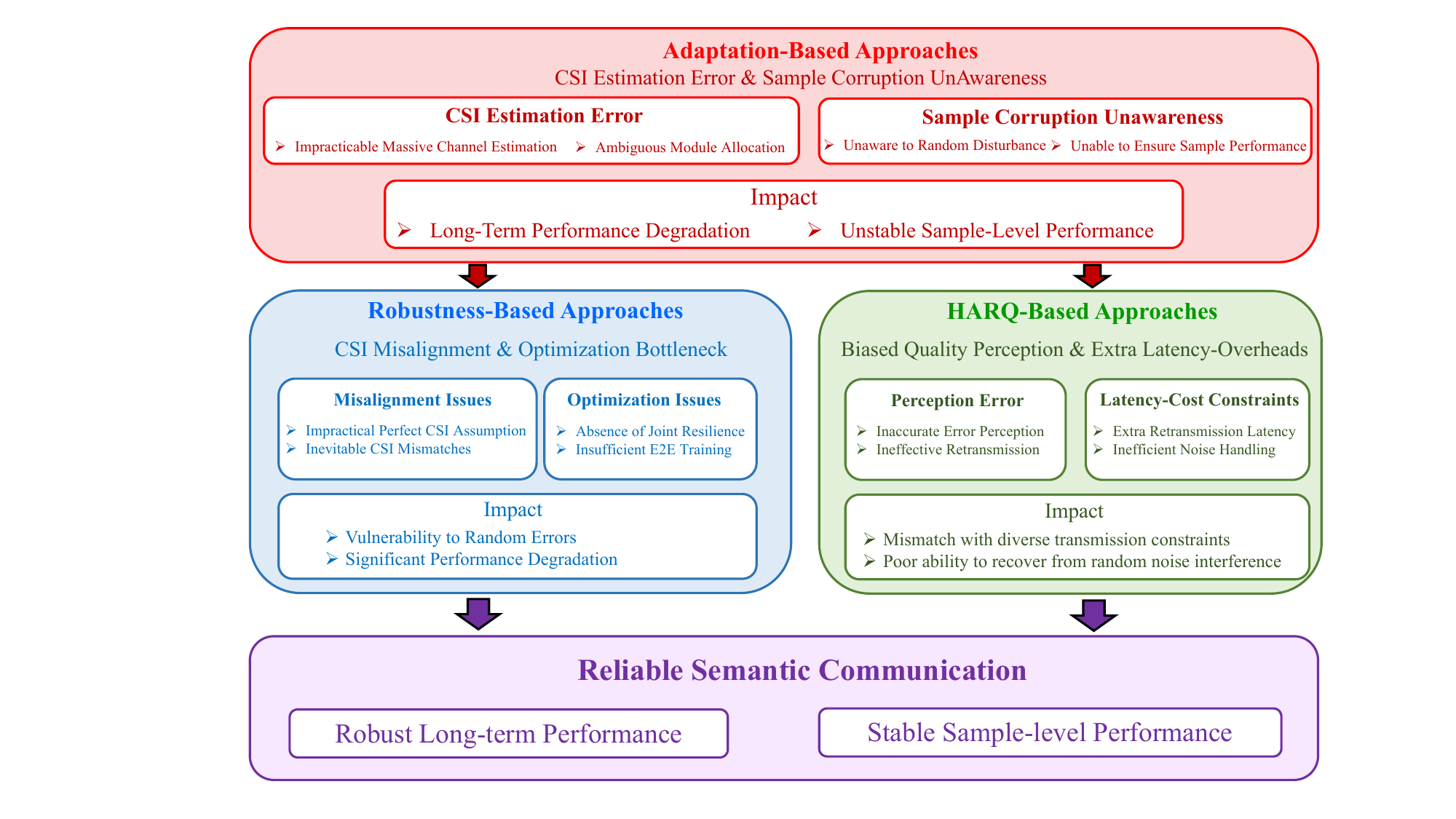}
    \caption{Key challenges in reliable semantic communication.}
    \label{fig:challenge}
\end{figure*}

\section{Conventional Approaches}
In this section, we discuss the key challenges of reliable semantic communication shown in Fig.~\ref{fig:challenge}, deducing some representative approaches to handling these challenges. 

\subsubsection{Adaptation-Based Approaches} 
Existing adaptation-based SemComm methods aim to improve robustness against channel variations by conditioning semantic encoding on CSI, most commonly the SNR. For example, \cite{10437864, CCHARQ} introduced rate-adaptation mechanisms that dynamically adjust the transmitted signal length according to channel conditions. Under unfavorable channels, these methods prioritize more important semantic features, thereby preserving task-relevant information under limited transmission resources. Instead of changing the codeword length, \cite{10589474, SCHARQ} proposed SNR-adaptive schemes that retain a fixed transmission length while reweighting semantic features according to side information. In this way, important features can receive stronger protection under different SNR conditions, improving the stability of semantic reconstruction or task inference.

The above studies mainly focus on point-to-point semantic transmission, where the encoder is adapted to a single receiver and a single channel condition. However, practical semantic communication systems may need to serve multiple users with heterogeneous channels and different downstream tasks. In such scenarios, a point-to-point adaptation strategy may fail to provide stable performance for all receivers. To address this issue, \cite{10738311} proposed a multi-branch compression and fusion architecture, where task-channel-aware sub-encoders extract user-specific semantic features and a fusion module integrates them into a unified broadcast signal. With an information bottleneck-based training strategy, this design improves robustness against channel and task heterogeneity, thereby enhancing reliable semantic transmission in multi-user broadcast environments.

Beyond scalar SNR adaptation, another line of work further exploits fine-grained CSI, such as channel coefficients, to support channel-aware semantic encoding in MIMO systems. For example, \cite{10510413, 10597355} incorporate channel matrix information into the semantic transmission process, enabling the model to better adapt feature mapping, power allocation, or feedback overhead to specific channel realizations. These studies demonstrate the potential of fine-grained CSI for improving channel adaptability and transmission reliability. For example, \cite{10510413} introduces semantic distortion outage probability, to characterize semantic reliability and adjusts CSI feedback overhead according to predicted image-level reconstruction quality, while DeepJSCC-MIMO proposed in \cite{10597355} uses channel heatmaps to adapt the JSCC model to MIMO channel conditions. To  guarantee each user's minimum reconstruction quality, \cite{hou2026reliable} develop a Bayesian optimization (BO)-based online algorithm that enables flexible control of the user-side semantic compression ratio (CR) and allocation of transmission rates.

However, adaptation-enhanced SemComm methods still face two major limitations from the perspective of reliable semantic communication. First, these methods are often bottlenecked by the assumption of accurately acquired CSI. In practical wireless systems, CSI may be noisy, delayed, quantized, or only partially observable. This issue becomes more severe when adaptation relies on fine-grained channel information, such as channel matrices, whose acquisition usually requires substantial pilot or feedback overhead and is vulnerable to estimation errors. As a result, adaptation mechanisms that are highly dependent on accurate CSI may suffer from significant performance degradation under CSI mismatch. Furthermore, this limitation becomes more pronounced in large-antenna MIMO systems\cite{10597355}, where fine-grained channel matrices can provide useful adaptation information but are difficult to acquire accurately due to pilot overhead, feedback cost, and channel estimation errors.

On the other hand, even when reasonably accurate average CSI is available, existing adaptation mechanisms are primarily tailored to optimize average transmission performance rather than ensuring sample-level reliability.
Due to the stochastic nature of channel noise, individual data samples may experience severe distortions that are not adequately captured by average quality metrics.
Thus, adaptation alone lacks the capacity to explicitly detect and recover from such anomalies, making it difficult to guarantee stable task performance under adverse channel conditions.

\subsubsection{Robustness-Based Approaches}
Beyond adaptation-based approaches, robustness-based methods aim to improve the intrinsic resilience of SemComm systems against random transmission errors and model-channel mismatch.
Different from channel-adaptive methods, which adjust semantic encoding according to the estimated channel state, robustness-based approaches seek to train a unified semantic codec that can maintain a stable performance across different channel realizations and noise conditions.
This is particularly important because practical transmission impairments are often stochastic and difficult to fully characterize before transmission.
Although \cite{10648086} avoids such impairments via pilot-free MIMO, the channel independent coding significantly limits transmission performance.
Accordingly it is necessary to enhance the robustness of SemComm systems for reliable transmission.

A representative direction is to enhance the robustness of the semantic codec during training.
For example, \cite{11178221} proposed an alternating multi-phase training strategy for digital semantic communication, where the training process consists of feature extraction, robustness enhancement, and training-testing alignment.
In the robustness enhancement phase, the codec is trained to tolerate information loss caused by quantization, modulation-demodulation, and bit-flipping effects.
Moreover, a mask-attack mechanism is introduced to simulate severe and unpredictable bit-flipping in a differentiable manner, while an information restoration network is employed at the receiver to recover corrupted semantic features before decoding.
In this way, the system can achieve stronger resilience against random digital transmission errors and support standard modulation schemes within existing digital communication frameworks.

However, robustness-based approaches still have inherent limitations from the perspective of reliable SemComm.
First, they mainly improve the prior noise resilience of the codec, but they do not explicitly identify whether a specific received sample has suffered severe distortion.
Second, robustness training usually improves the average tolerance to random errors, but may still fail to protect lower-tail samples under rare but severe channel perturbations.
Therefore, although robustness-oriented codec design is essential for reliable semantic communication, it is not sufficient by itself to guarantee stable sample-level performance under highly uncertain transmission conditions.

\subsubsection{HARQ-Based Approaches}
Compared with adaptation- and robustness-based approaches, HARQ-based methods are not primarily codec-level designs.
Instead, they introduce a higher-level reliability-control mechanism after transmission based on receiver-side feedback.
The key advantage is that the receiver can evaluate the realized quality of each received sample, and thus infer whether the current channel realization and stochastic noise have caused severe semantic distortion.

By introducing perceptual or task-oriented quality metrics at the receiver, HARQ-based methods can trigger selective retransmission or incremental semantic refinement for unreliable samples.
More specifically, by designing fine-grained error awareness mechanisms that estimate noise-induced semantic distortion, these methods can prioritize the features that have the greatest impact on reconstruction quality or task performance. For example, \cite{IK} models the transmission reliability of SemComm as the amount of intact information available for downstream task. By performing an incremental data retransmission for error correction, it enables a balance between transmission efficiency and data reliability.  As another classic retransmission strategy, \cite{BiHARQ} asks for identical retransmission based on current SNR condition as well as task performance. By replacing corrupted semantic features by retransmitted feature, it performs higher classification performance under diverse channel conditions. Moreover, \cite{S3CHARQ} treats semantic features and check codewords as complementary information extracted from the original image in multi-round transmission scenarios. By jointly optimizing the feature extraction and check coding, it maximizes communication efficiency under stringent communication overheads and effectively enhances transmission reliability.
These attempts allow the system to extract complementary information from retransmitted features and achieve more effective error compensation.
Therefore, HARQ-based approaches provide a promising way to improve lower-tail reliability and stabilize task performance under adverse channel conditions.

However, the reliability gain of HARQ-based approaches is obtained at the cost of additional communication and latency overhead, including  {acknowledgment (ACK) and negative acknowledgment (NACK) feedback}, extra retransmission resources, and increased round-trip time.
Moreover, their effectiveness depends heavily on the accuracy of receiver-side quality estimation and the efficiency of retransmission decision making.
If the distortion estimator fails to identify severely corrupted samples, or if the retransmission policy is not well matched to the task requirement, the additional feedback and retransmission may lead to inefficient resource usage.

\section{Remaining Gaps and Design Implications}
The above discussions show that adaptation-based, robustness-based, and HARQ-based approaches address reliability from different but complementary perspectives. 
Adaptation-based methods adjust the semantic representation or transmission policy according to time-varying channel conditions. 
However, most adaptation mechanisms are designed to improve expected performance under estimated channel states, rather than to explicitly protect samples that suffer severe distortion under unfavorable instantaneous channel realizations.
Robustness-based methods improve the intrinsic resilience of the semantic codec against stochastic transmission errors and model-channel mismatch. 
However, they are unable to directly observe whether a particular received sample has been corrupted after transmission, and the robustness-oriented training may introduce a tradeoff between average performance and reliability, which is insufficient to guarantee lower-tail reliability under rare but harmful channel perturbations.
HARQ-based methods can trigger retransmission or refinement when severe semantic distortion is detected. 
However, its reliability gain is obtained at the cost of additional feedback, communication overheads, and round-trip latency. 
These limitations indicate that reliable SemComm should not rely on a single mechanism. 

Instead, future designs need to combine the strengths of different reliability strategies. 
First, adaptation should be integrated with robustness-oriented training, so that channel-aware semantic encoding can remain effective even under imperfect CSI. 
This motivates robust adaptive designs such as the knowledge distillation-based solution discussed in Section \ref{sec:solution}, which jointly considers channel-matrix adaptation and resilience to CSI estimation errors. 
Second, retransmission should be coupled with semantic feature design and receiver-side distortion perception, so that additional transmissions can provide complementary information rather than simply repeating corrupted content. 
This motivates the HARQ-oriented solution in Section \ref{sec:solution}, where feature coding, check coding, and adaptive retransmission are jointly considered to improve sample-level reliability. 

\section{Potential Solutions}
\label{sec:solution}
\begin{figure}[!t]
    \centering
    \subfloat[The overview of HANA-JSCC.]{
        \includegraphics[width=0.85\linewidth, trim=0cm 0cm 0cm 0cm,clip]{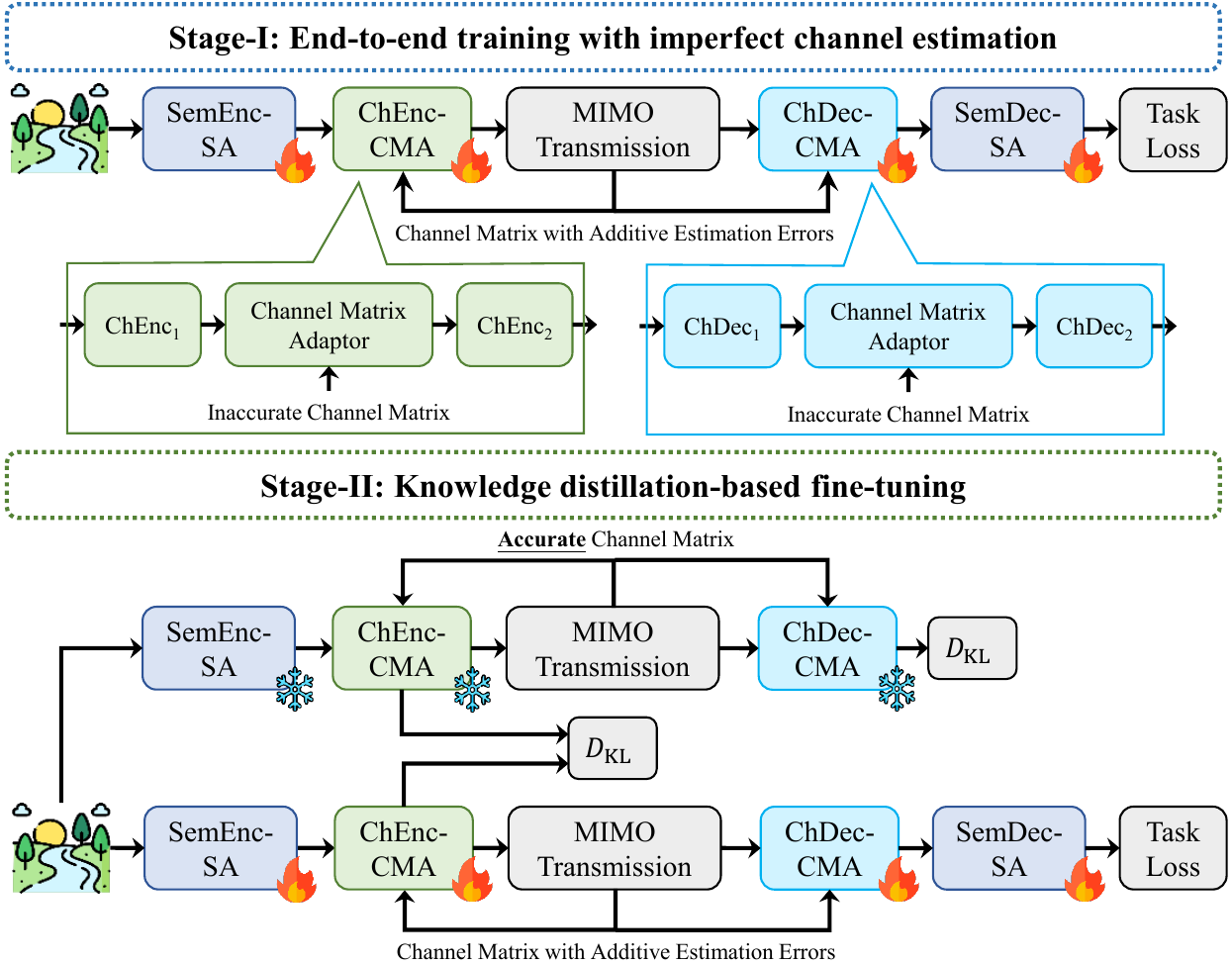}%
        \label{fig:HANAJSCC_OVERVIEW}} \\ 
    \subfloat[Simulation results of HANA-JSCC and benchmarks.]{
        \includegraphics[width=0.82\linewidth, trim=0cm 0cm 0cm 0cm,clip]{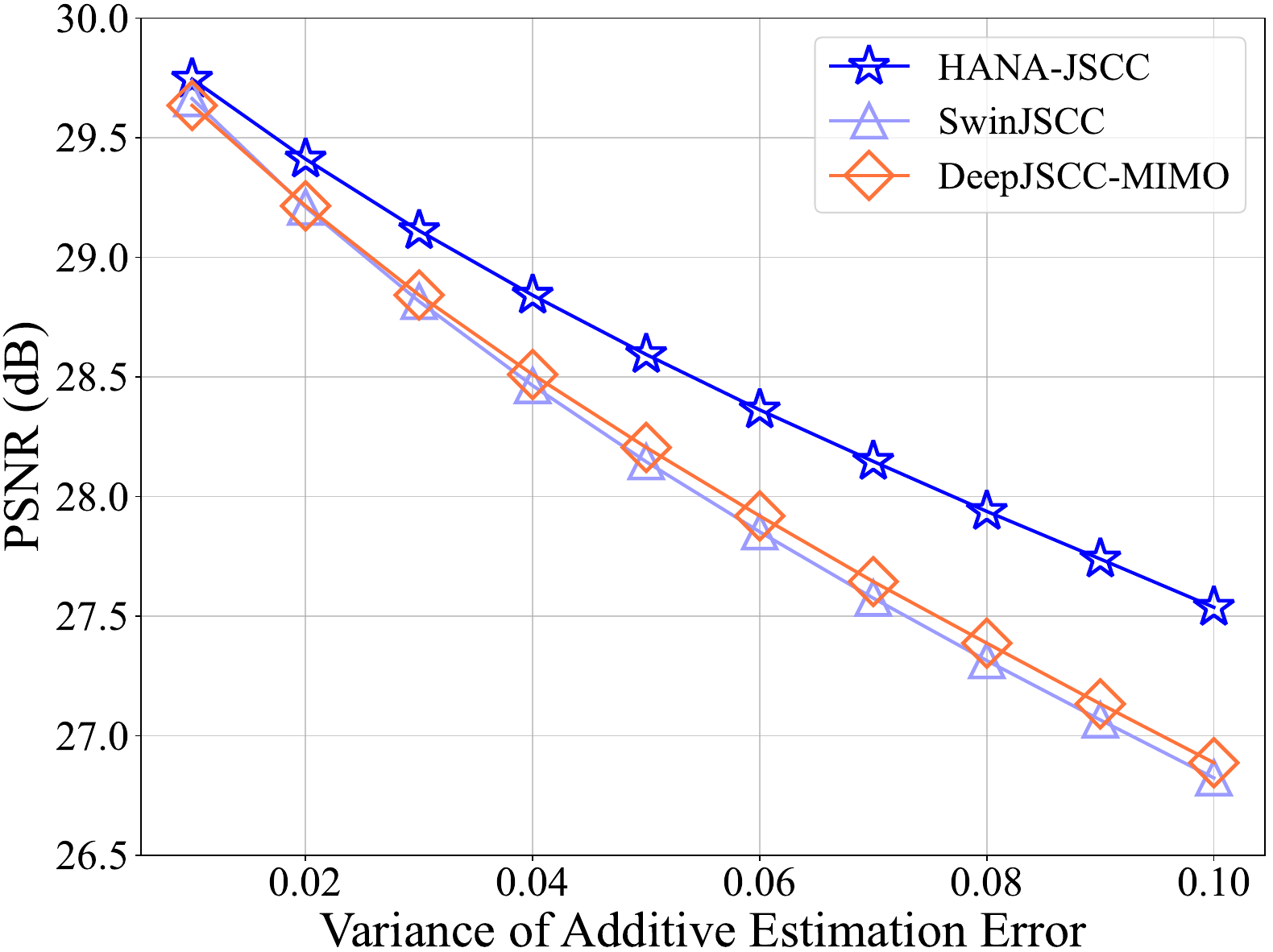}%
        \label{fig:HANAJSCC_RESULTS}
        }
    \caption{Framework of HANA-JSCC and evaluation results on CIFAR10 dataset. }
    \label{fig:HANAJSCC}
\end{figure}  

\subsubsection{Boosting robust SemComm with imperfect CSI via knowledge distillation} 
Inspired by SNR-adaptive JSCC schemes, several studies have leveraged CSI feedback in MIMO systems to enhance robustness against fading effects during transmission. 
However, while SNR estimation is relatively straightforward, accurately acquiring the channel gain (i.e., the channel matrix) remains a formidable challenge. 
Furthermore, obtaining perfect full channel matrices in practical MIMO systems is both costly and difficult due to stringent constraints on pilot overhead and hardware resources, particularly in large-scale MIMO scenarios. 
Consequently, ensuring reliable communication necessitates enhancing the system's robustness against estimation errors. 
To address this issue, as shown in Fig.~\ref{fig:HANAJSCC_OVERVIEW}, \cite{11372929}  {proposed HANA-JSCC}, an approach refining the compressed latent semantics using the acquired channel matrix during the channel coding phase, thereby improving robustness against diverse channel realizations. 
Additionally, \cite{11372929} developed a knowledge distillation-based training strategy to further augment the system's resilience against channel matrix estimation errors. 
By emulating a perfect-CSI teacher model, this approach effectively mitigates the performance degradation caused by channel matrix mismatches.

In the following, we evaluate HANA-JSCC against two benchmarks: SwinJSCC \cite{10589474} and DeepJSCC-MIMO \cite{10597355}. 
Following \cite{11372929}, practical channel estimation errors are simulated by injecting zero-mean Gaussian additive noise into the channel matrix across varying SNRs.

Fig.~\ref{fig:HANAJSCC_RESULTS} compares the image reconstruction quality, measured in PSNR, across the three evaluated schemes. 
Overall, HANA-JSCC consistently delivers superior performance across a wide range of SNRs and estimation error levels, showcasing its distinct advantage under misaligned channel conditions. 
On average, it maintains a noticeable PSNR lead in image reconstruction over both DeepJSCC-MIMO and SwinJSCC. 
In a nutshell, these results validate that the HANA-JSCC effectively achieves robustness against channel estimation errors.

\begin{figure}[!t]
    \centering
    \subfloat[The overview of S3CHARQ.]{
        \includegraphics[width=0.85\linewidth, trim=5cm 2cm 5cm 1cm,clip]{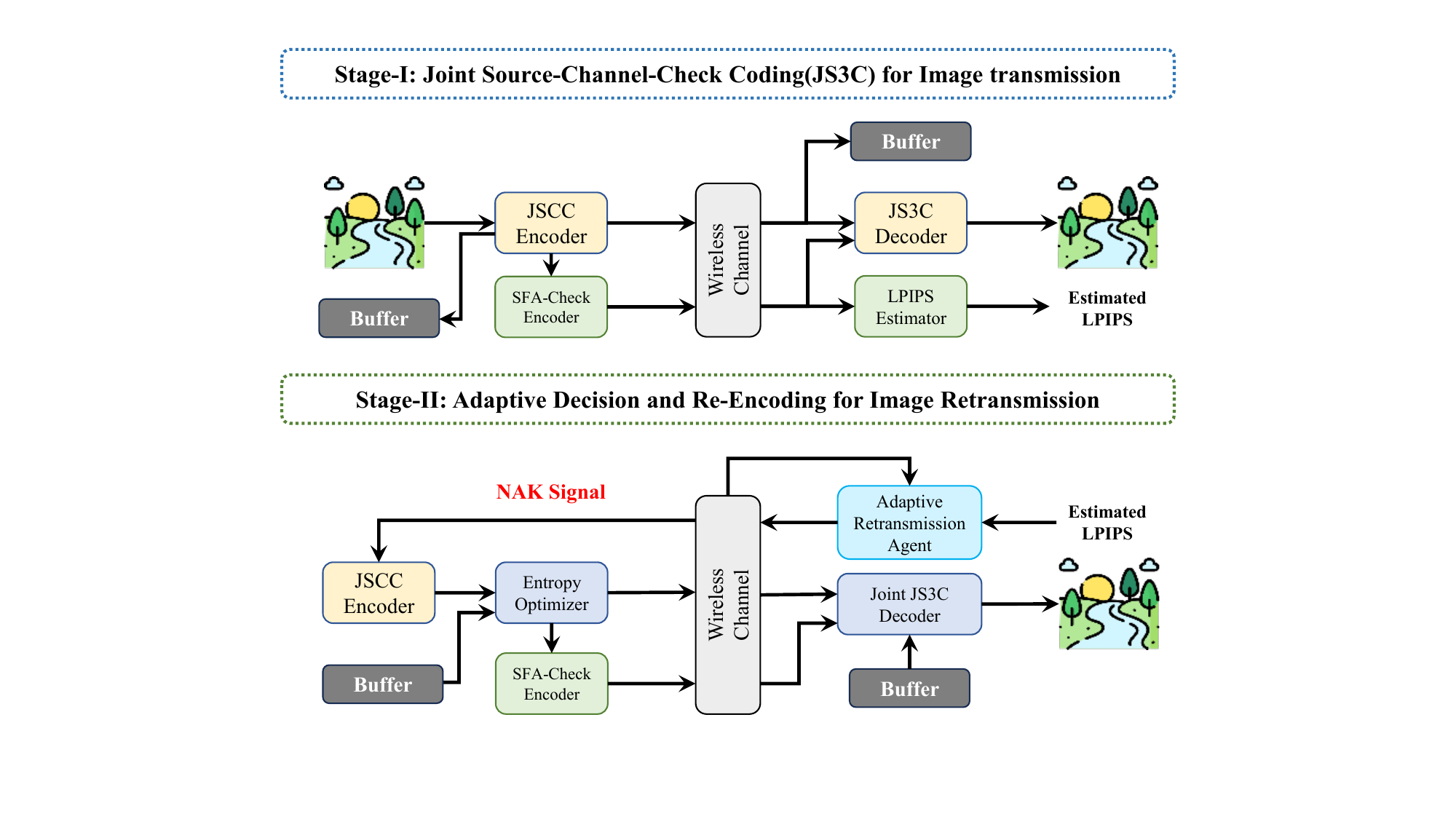}%
        \label{fig:S3CHARQ_OVERVIEW}} \\ 
    \subfloat[Simulation results of S3CHARQ and benchmarks.]{
        \includegraphics[width=0.82\linewidth, trim=0cm 0cm 0cm 0cm,clip]{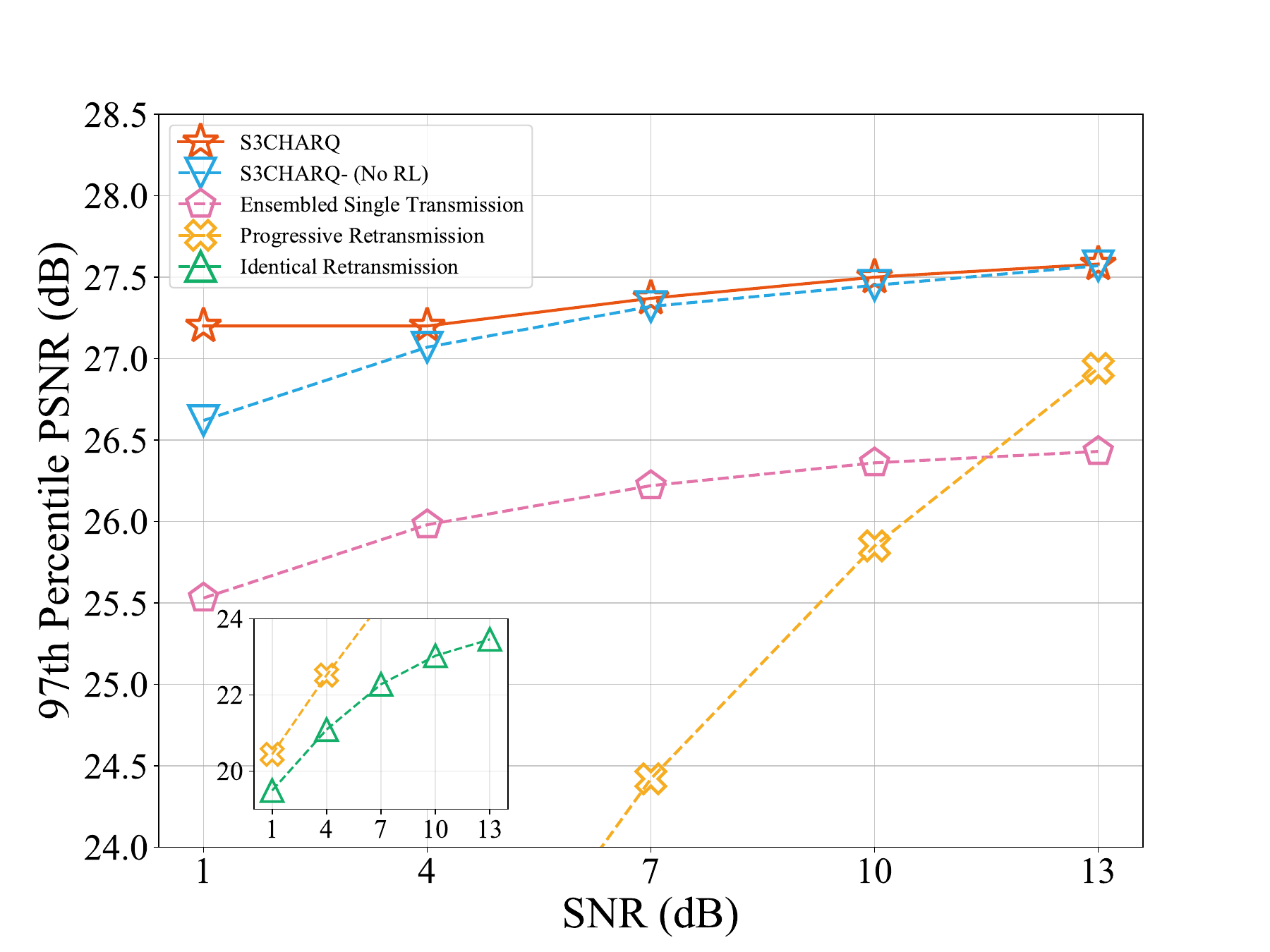}%
        \label{fig:S3CHARQ_RESULTS}
        }
    \caption{Framework of S3CHARQ and evaluation results on CIFAR10 dataset. }
    \label{fig:S3CHARQ}
\end{figure}  

\subsubsection{Joint Optimization of Feature Coding, Check Coding, and Retransmission} To ensure high-fidelity image reconstruction, recent efforts have explored incremental transmission of check codewords, where the receiver estimates perceptual quality and triggers additional semantic transmission for samples falling below a predefined threshold.
While such designs improve the system performance lower bound, they typically treat feature coding and check-coding as independent procedures, overlooking the fact that check codewords inherently carry latent semantic information. Moreover, the effectiveness of retransmission is fundamentally constrained by quality estimation errors, leading to suboptimal identification of severely corrupted samples. Additionally, homogeneous retransmission features often provide limited information diversity, resulting in insufficient compensation for accumulated noise distortion. To address these limitations, a joint source-channel-check coding retransmission framework, namely S3CHARQ, has been proposed in \cite{S3CHARQ}. As shown in Fig.~\ref{fig:S3CHARQ_OVERVIEW}, under a unified information-theoretic perspective, the framework shapes the information distribution between semantic features and check codewords. By enabling the latter to provide complementary information while preserving perceptual fidelity, it achieves enhanced reconstruction quality by jointly decoding received features and check codewords. Furthermore, to mitigate the mismatch between imperfect quality estimation and threshold-based decision, S3CHARQ incorporates a reinforcement learning-based adaptive retransmission module, which dynamically determines retransmission actions based on channel conditions and sample characteristics. In the retransmission phase, the system performs feature re-encoding conditioned on prior transmission, thereby provides a more precise and efficient reliability enhancement mechanism, effectively guaranteeing the system performance lower bound under adverse channel conditions.

Subsequently, we evaluate the performance of S3CHARQ on the CIFAR-10 dataset under diverse SNR conditions, and compare it with several representative schemes, i.e., single transmission with aligned communication overheads\cite{10589474}, progressive retransmission\cite{CCHARQ}, as well as identical retransmission\cite{SCHARQ}.
To further isolate the contribution of joint source-channel-check coding and adaptive retransmission decision, we additionally consider a threshold-based variant S3CHARQ(No RL), where the retransmission decisions are configured with a compression ratio and retransmission ratio of 1/8, and transmissions are conducted over AWGN channel.
Moreover, to eliminate the impact of threshold selection on performance, we followed alignment strategy of \cite{S3CHARQ} across all schemes.

Fig.~\ref{fig:S3CHARQ} presents a comparative analysis in terms of both average PSNR and tail PSNR, a.k.a the 97th percentile PSNR. The latter characterizes the system performance lower bound under unfavorable conditions.
Among all evaluated retransmission schemes, S3CHARQ consistently achieves the best average PSNR performance across all SNR regimes. More importantly, it also demonstrates a clear advantage in tail PSNR, indicating its superior capability in safeguarding performance under severe distortion.
These results validate the effectiveness of joint source-channel-check coding and retransmission-aware re-encoding in simultaneously improving average performance and tightening the system performance lower bound.
In addition, compared to its static-threshold counterpart, the proposed S3CHARQ exhibits consistent performance gains.
This further highlights the importance of accurate and adaptive retransmission decisions, which enable more precise recovery of severely corrupted samples.
As a result, the proposed framework effectively enhances both data fidelity and transmission reliability for image reconstruction tasks.

\section{Future Research Directions}
Building on current progress, future research on reliable semantic communication should focus on more rigorous reliability metrics, better compatibility with practical wireless systems, and deeper joint design of robustness and retransmission.
\subsection{Joint Design of Robustness and Retransmission}
A particularly promising direction is the deeper integration of robustness-oriented encoding and retransmission-oriented transmission control. At present, these two lines of research are often treated separately: robustness methods mainly improve resistance to channel uncertainty during the initial transmission, while retransmission methods enhance reliability after receiver-side quality estimation. However, these two mechanisms are inherently coupled. Future work should study unified designs in which semantic feature protection, distortion perception, retransmission triggering, and re-encoding are jointly optimized. Such an integrated framework may provide a more effective way to improve not only average performance, but also lower-tail reliability in highly dynamic and uncertain environments.

\subsection{Reliability Metrics Beyond Averages}
Another important direction is to establish more rigorous reliability metrics for semantic communication. Existing works usually report average PSNR, average task accuracy, or mean perceptual quality, but these metrics cannot fully reflect system behavior under adverse transmission conditions. For reliable SemComm, it is important to go beyond average-case evaluation and explicitly characterize lower-tail performance, outage behavior, and task failure risk. In particular, for JSCC-based semantic systems, a clearer and more principled reliability definition is still needed, so that robustness under rare but severe channel degradations can be evaluated in a unified and comparable manner.

\subsection{Compatibility with Digital Wireless Networks}
A third important direction is to improve compatibility with practical communication systems. Most current reliable SemComm designs are developed in an end-to-end analog JSCC manner, which offers strong performance potential but is not always easy to integrate into existing digital wireless architectures. Future research should therefore investigate how reliable semantic communication can coexist with digital semantic communication frameworks and with current protocol stacks. This includes modular design, flexible redundancy control, and smooth deployment over existing network infrastructure. Such compatibility will be important for moving reliable SemComm from proof-of-concept algorithms to practical systems.

\section{Conclusion}

In this article, we discussed reliable semantic communication beyond average performance. In particular, we reviewed adaptation-based, robustness-based, and HARQ-based approaches, and clarified their complementary roles and limitations. We further introduced two representative solutions, namely robust adaptive semantic communication under imperfect CSI and joint source-channel-check coding with adaptive retransmission, to demonstrate how reliability can be enhanced from both codec-level and feedback-driven perspectives. Finally, we highlighted several future research directions, including joint robustness-retransmission design, reliability metrics beyond averages, and compatibility with digital wireless networks. This article provides insights for developing reliable SemComm systems with stable performance under dynamic and uncertain environments.
\bibliography{ref}    
\bibliographystyle{ieeetr}
\end{document}